\documentclass[reprint,
 amsmath,amssymb,
 aps,
]{revtex4-2}
\usepackage{graphicx}
\usepackage{dcolumn}
\usepackage{hyperref}
\usepackage{bm}

\begin{document}

\title{Reliable Detection of Causal Asymmetries in Dynamical Systems}

\author{Erik Laminski}
\email{erik@neuro.uni-bremen.de}
\author{Klaus R. Pawelzik}%
\email{pawelzik@neuro.uni-bremen.de}
\affiliation{University of Bremen, 28359 Bremen, Germany}

\date{\today}

\begin{abstract}
Knowledge about existence, strength, and dominant direction of causal influences is of paramount importance for understanding complex systems. With limited amounts of realistic data, however, current methods for investigating causal links among different observables from dynamical systems suffer from ambiguous results. Particularly challenging are synchronizations, where it is difficult to infer the dominant direction of influence. Missing is a statistically well defined approach that avoids false positive detections while being sensitive for weak interactions. The proposed method exploits local inflations of manifolds to obtain estimates of upper bounds on the information loss among state reconstructions from two observables. It comes with a test for the absence of causal influences. Simulated data demonstrate that it is robust to intrinsic noise, copes with synchronizations, and tolerates also measurement noise. 
\end{abstract}


\maketitle

\section{Introduction}
Simple cause-effect notions of causality are misleading when interactions are reciprocal. This is particularly clear for deterministic dynamical systems where Takens' Theorem \citep{Takens_1981, Packard_1980, Stark_1997} shows that different observables individually contain all information about the entire system's state. Here, state reconstructions from different observables are generically equivalent which reflects the fact that the systems are non-separable and behave as wholes. Still, it is of considerable interest to gain knowledge about the strength of influences among selected components of a complex dynamical system. 

Several methods proposed for this purpose rely on phase space reconstructions \citep{Sugihara_2012, Harnack_2017, Benko_2018}. In essence, they are based on the following simple consideration: Let a system X uni-directionally influence another system Y. Obviously, then  Y receives information about X. Consequently, states of Y will contain information about the state of X, while states of X by assumption cannot provide full 'knowledge' about Y. Thereby states Y can be expected to predict observables of X better than vice versa. This heuristics was put on solid mathematical grounds in \citep{Harnack_2017} leading to a method that allowed to capture also mutual interactions. The measure of Topological Causality was introduced for inference of the intensity of directed effective influences among observables. It relies on local expansions of the mappings \textit{between} state reconstructions. While mathematically transparent, Topological Causality was hitherto estimated from fitting the local maps among state reconstructions from data. This can be quite challenging, in particular, when only limited amounts of observations are available, when the system's strict determinism is violated (intrinsic noise), and when measurement noise contaminates the data. Particularly, synchronizations and common input can cause misleading results. Similar problems affect also  other methods based on the same heuristics. E.g. the CCM method \citep{Sugihara_2012} was found to yield wrong directions of dominant influences with synchronizations \citep{Deyle_2016,Baskerville_2017}, particularly see Fig.4 in \citep{Krakovska_2016}. 

The method presented here is also based on expansions, however, not \textit{among} but instead \textit{within} each state space reconstruction. Following the above heuristics, it estimates a bound on the relative amount of information contained in the state reconstructed from of the causally affected system about the state reconstructed from the effecting system. This is done by comparing the sizes of local neighbourhoods in one state space reconstruction with both, the projections of neighbours in the other state space reconstruction and random neighbours. The corresponding manifold inflations not only capture possible dimensional conflicts among state reconstructions \citep{Benko_2018}, but also allow for a statistical criterion to control for false positive detections of causal influences. 

After introducing the cross projection method (CPM) it is tested on a time discrete model system demonstrating the relation to coupling constants and its dependency on intrinsic noise, limited measurements, and measurement noise. Then we demonstrate that CPM reliably detects the absence of causal links and copes with synchronization when applied to time series of realistic length from time continuous systems. A first application to heart rate and breathing rate data is found to deliver unambiguous results. 

\section{Manifold inflation as a proxy for information loss}
We consider dynamical systems composed of two subsystems $X$ and $Y$, governed by 
\begin{align*}
\mathbf{\dot{x}} = f(\mathbf{x}, w_{xy}\mu_x(\mathbf{y})) \\
\mathbf{\dot{y}} = g(\mathbf{y}, w_{yx}\mu_y(\mathbf{x}))
\end{align*}
where $\mu_i(\mathbf{i})$ denote fixed scalar functions and $w_{ij}$ coupling constants. It was shown by Takens \citep{Takens_1981} that a topologically equivalent portrait of the attractor can be reconstructed from lagged coordinates. For this purpose delayed copies of a single observable are merged giving a $m$-dimensional vector $\mathbf{r}^x(t)=[\mathbf{x}(t), ...,\mathbf{x}(t+(m-1)\tau)]$. The dimension $m$ is sufficient if $m > 2D$ where $D$ is the dimension of the attractor. Formally the time delay $\tau > 0$ is arbitrary, in practice, however,  there are methods for finding suitable values of $m$ and $\tau$ \citep{Liebert_1991}\citep{Grassberger_1991}. Note that to be a valid reconstruction of the overall attractor, there must be incoming connections $w_{xi}$ linking the subsystem X to the whole system. In this case there also exists a unique mapping between reconstructions from different observables, in this case from $\mathbf{r}^x$ to $\mathbf{r}^y$ \citep{Gedeon_2015}. \\

\begin{figure}[h!]
\includegraphics[width=0.5\textwidth]{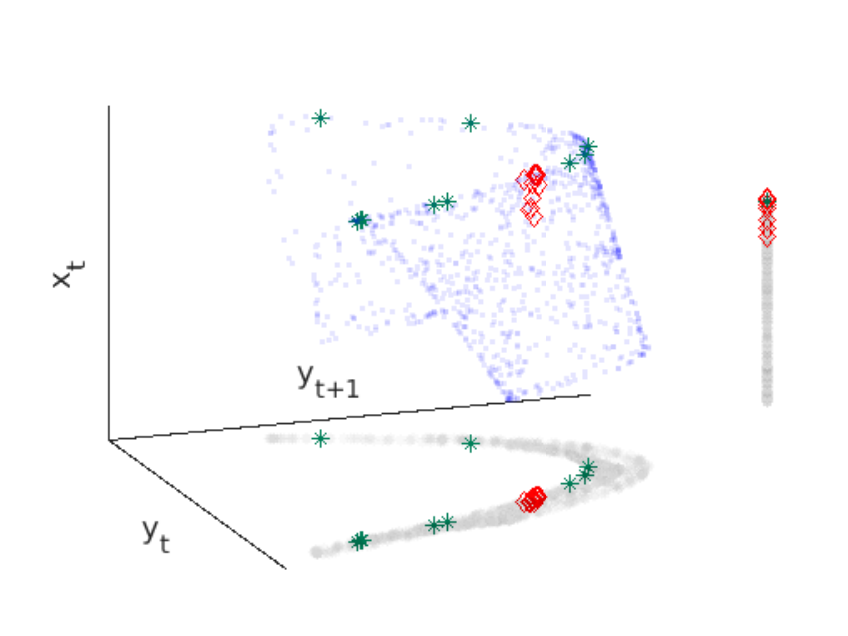}
\caption{$x_t$ over $y_t$ and $y_{t+1}$ for the noise free unilaterally coupled logistic maps with $w_{xy}=0,w_{yx}=0.3, R_x=R_y=3.82$, together with the projection of the manifold on the $y_{t+1}$-$y_t$-plane and the $x_t$-axis. $10^3$ data points are shown in blue the 10 nearest neighbours of a reference point are shown in colour in both, the manifold and the projections (grey). The 10 neighbours searched in Y (X) are shown in red (green).}
\label{fig:figure1}
\end{figure}

Here we use a combination of local and global properties of the relations among nearest neighbours to reference points in both reconstructions for estimating how much the causally affected system 'knows' about the state of the system influencing it. Each point is identified by its time index $t$ and its location in the respective reconstruction. For each reference point $t$ we determine the $k$ nearest neighbours in both reconstructed spaces $t_l^x$ and $t_l^y, l = 1,...,k$ and the (euclidean) distances from the reference point to these neighbours both, in their origin space $L_x(t,t_l^x)$ and the distance to the putative neighbours based on proximity in the respective other space $L_x(t,t_l^y)$.\\
For the chance level an ensemble $E$ of surrogate neighbors is generated by shifting the time indices of all actual neighbors of each reference point with a mutual random time interval $\delta t_E$
We avoid overlap of the embedding vector by excluding the immediate temporal neighbors (i.e. $\delta t_r > \pm 2m\tau$). When applying this random time shift to the time indices of neighbors in space $Y$ a given reference point $t$ becomes in $X$ associated with a set of points $t_l^{y^*}$ haveing distances $L_x(t,t_l^{y^*})$ to these points. This procedure removes causal relations of neighborhoods between different observables while preserving the temporal correlations within the set of surrogate neighbors. \\
 Also, we introduce $d_i^j(k)$ as mean logarithmic size of the neighbourhood for all $L_i(t,t_l^j)$:
\begin{align*}
d^j_i(k) =  \left< log(max[(L_i(t,t_l^j)]_{l=1..k}) \right>_{t \land E} \quad i,j= x,y,x^*,y^*
\end{align*}
The random neighbourhoods are here averaged over both, the Monte-Carlo ensembles $E$ and the reference points, whereas for the non-shifted neighbours the average is only over the reference points. \\
As an example consider two coupled logistic maps
\begin{align*}
x(t+1) = x(t)[R_x (1-x(t)) - w_{xy}y(t)] + \eta_x(t)\\
y(t+1) = y(t)[R_y (1-y(t)) - w_{yx}x(t)] + \eta_y(t)
\end{align*}
with reflecting boundaries, $R_i$ being system parameters and subjected to additive Gaussian noise $\eta_i(t) \in \mathcal{N}(0,\sigma)$ . The simplest case is the noise free unilaterally coupled system ($w_{xy}=0$ and $w_{yx}=0.3$). Here, the time-delay reconstruction of subsystem X is one-dimensional and we can therefore visualize the whole system in three dimensions by showing $x_t$ over the $(y_t,y_{t+1})$-plane. (FIG. \ref{fig:figure1}).
Since $w_{yx}>0$ information about X is contained in Y and the images of neighbours in Y are also localized in X. With increasing coupling $w_{yx}$ neighbours searched in Y become more localized in X, in the limiting case of perfect information preservation the neighbours are identical and $d_x^x(k)=d_x^y(k)$. In contrast, X here does not constrain Y completely and thus the image of these neighbours is spread over the whole $y_t$-$y_{t+1}$-plane.  In the other limiting case, $w_{xy}=0$, no information about X is included in Y and $d_x^y(k)$ will on average be identical with the random neighbourhood $d^{y^*}_x(k)$. \\

These relations between neighborhood sizes are visible in plots of the mean logarithmic neighbourhood sizes $d_i^j(k)$ as functions of $\kappa=\psi(k)-log(N)$, where $k$ is the number of neighbours and $N$ the amount of data and $\psi$ is the digamma-function. This particular choice of the abscissa allows for an unbiased estimate of the fractal (information) dimensions $D_1$ of the subsystems by taking the slope of $\psi(k)$  versus $d_x^x(k)$ (resp. $d_y^y(k)$) \citep{Grassberger_1988}. In the present example the manifold reconstructed from observable X has a smaller dimension than the attractor reconstructed from the influenced observable Y. While dimensional conflicts provide a sufficient criterion for the direction of causal influence in unilaterally coupled deterministic systems \citep{Benko_2018}, they are useless for mutually coupled systems. In this general case a different criterion for determining the dominant direction of causal influence is needed. 
 
\begin{figure}[h!]
\includegraphics[width=0.5\textwidth]{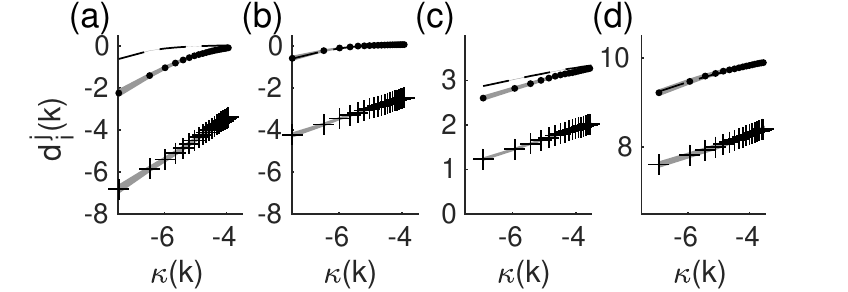}
\caption{Logarithmic neighbourhood sizes $d_i^j(k)$ over $\kappa$ for noise free unilateral coupled logistic maps ($w_{xy}=0, w_{yx}=0.2$ and $R_x=R_y = 3.82$) from $N=10^3$ data points. The observables were embedded with $m=4$ and $\tau=1$ and $10^2$ 
 ensembles were used for chance-level-estimation. \textbf{a)} $d_x^x(k)$ shown as solid line, $d_x^y(k)$ shown as dotted line and the respective chance-level $d_x^{y^*}(k)$ (dashed line). \textbf{b)} $d_y^y(k)$ shown as solid line, $d_y^x(k)$ shown as dotted line and the respective chance-level $d_y^{x^*}(k)$ (dashed line). Furthermore the fivefold standard error is shown for each $d_i^j(k)$ as a grey shade.\\
\textbf{c)}, \textbf{d)} show the same results, e.g. neighborhoodsizes, for the heart and breathing rate of a sleeping human. The data was sampled at $2Hz$ and embedded with $\tau=6$ and $m=5$, all $n=601$ data points were chosen as reference points. \textbf{c)} $d_{Heart}^{Heart}(k)$ shown as solid line, $d_{Heart}^{Breath}(k)$ shown as dotted line and the respective chance-level $d_{Heart}^{Breath^*}(k)$ (dashed line). Due to the small amount of data only the threefold standard deviation was used to determine the chance-level. \textbf{d)} $d_{Breath}^{Breath}(k)$ shown as solid line, $d_{Breath}^{Heart}(k)$ shown as dotted line and the respective chance-level $d_{Breath}^{Heart^*}(k)$ (dashed line). }
\label{fig:figure2}
\end{figure}

With $d_x^x(k)$ providing a lower bound and $d_x^*(k)$ an upper bound for the size of the $k$-the neighbourhood, the size $d_y^x(k)$ can be used to define a measure for the information preserved within the neighbourhood in $y$.  We use the ratio of the distance between $d^x_x(k)$ and $d^y_x(k)$ and the chance-level $d^{y^*}_x(k)$:
\begin{align*}
I_{x\rightarrow y}(k) = \frac{d_x^{y*}(k) - d_x^y(k)}{d_x^{y^*}(k) - d_x^x(k)}
\end{align*}
Note that $0 \leq I_{x\rightarrow y} \leq 1$ and therefore $I_{x\rightarrow y} \simeq 0$ means that $x$ does not influence $Y$ at all. If $I_{x\rightarrow y} \simeq 1$ $Y$ 'knows everything' about $X$, which suggests that $X$ has a strong influence on $Y$. 
Furthermore we introduce a measure for the asymmetry of causal influences $\alpha$:
\begin{align*}
\alpha(k) = \frac{I_{y\rightarrow x}(k) - I_{x\rightarrow y}(k)}{I_{y\rightarrow x}(k) + I_{x\rightarrow y}(k)}
\end{align*} 
For determining significance we use the standard error (SE) of the mean logarithmic neighbourhood sizes $\sigma_{\bar{d_i^j}}$. For significance the standard errors $\sigma_{\bar{d_x^y}}$ and $\sigma_{\bar{d_x^*}}$ must not overlap, e.g. as shown in FIG. \ref{fig:figure2}  \textbf{a)}. In practice the errors overlap for large $k$. Therefore, in all following results we require for significance that the difference exceeds several SE's and additonally that at least 15 out of $k=1..20$ values do not overlap.

\section{Results}
\begin{figure}[h]
\includegraphics[width=.5\textwidth]{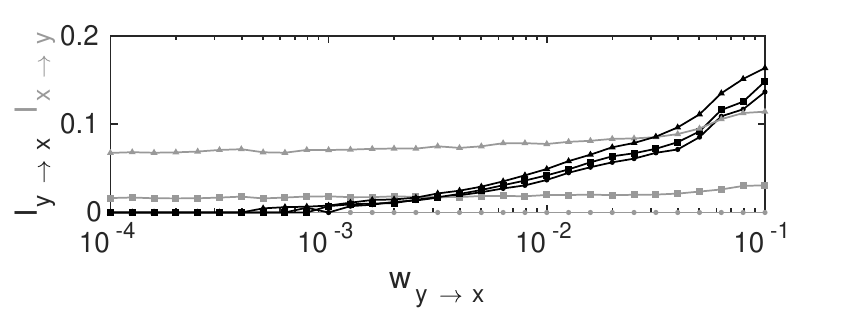}
\caption{Causal Influence between bilaterally coupled logistic maps. $2\cdot10^4$ data points were generated with $R_x=R_y=3.92$, the observed time series were embedded with $m=4$ and $\tau=1$ and we used the $k=20$ nearest neighbors to determine the causal influence and $E=10^3$ permutations for the chance-level. Black lines show $I_{y \to x}$ corresponding to the varied coupling $w_{y \to x}$, grey lines show the reverse direction with the coupling $W_{x \to y}$ fixed at $0$ (circles), $0.00316$ (squares) and $0.032$ (triangles).}
\label{fig:figure3}
\end{figure}

\begin{figure}[h!]
\includegraphics[width=.5\textwidth]{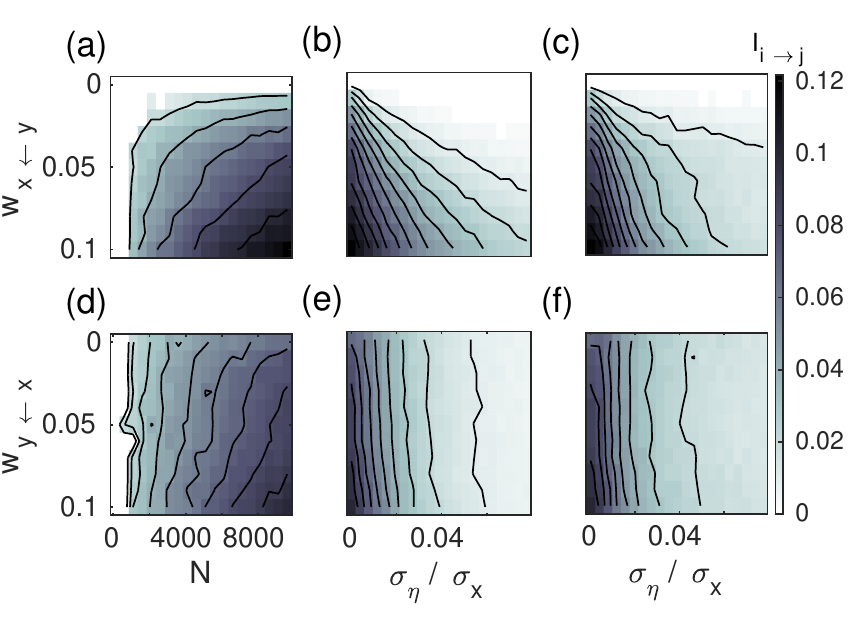}
\caption{Bilaterally coupled logistic maps $R_x=R_y=3.92$, $w_{x \to y} = 0.05$ and varying $w_{y \to x}$ between $0$ and $0.1$. All time series were embedded with $\tau=1$ and $m=4$ and $10^3$ ensembles were generated to estimate chance-level. \textbf{(a)} \& \textbf{(d}) Varying amount of data between $10^3$ and $10^4$ using $E=10^3$ permutations for chance-level, in \textbf{(a)} the causal influence $I_{y \rightarrow x}$ associated with the coupling $w_{yx}$ and \textbf{(d)}  the reverse direction $I_{y \rightarrow x}$. For noise polluted time-series $10^4$ data points were used. \textbf{(b)} \& \textbf{(e}) Additive internal noise is injected into the system. The $x$-axis shows the ratio of the standard deviations of noise and the unpolluted system varying between $0$ and roughly $8\%$ noise. \textbf{(b)} shows the causal influence $I_{y \rightarrow x}$ associated with the coupling $w_{yx}$ and \textbf{(e)} the reverse direction $I_{y \rightarrow x}$. \textbf{(c)} \& \textbf{(f)} Additive external noise is added to the observed time-series. The $x$-axis shows the ratio of the standard deviations of noise and the unpolluted system varying between $0$ and roughly $8\%$ noise. \textbf{(c)} shows the causal influence $I_{y \rightarrow x}$ associated with the coupling $w_{yx}$ and \textbf{f} the reverse direction $I_{y \rightarrow x}$. For better visualisation contour lines mark lines of equal causal influence.}
\label{fig:figure4}
\end{figure}

Firstly, we investigate sensitivity and specificity of the method. For this purpose two logistic maps are coupled bilaterally and the causal influence is determined for different coupling weights as shown in Fig. \ref{fig:figure3}. The causal influence $I_{y \to x}$ is a monotonic function of the weight $w_{y \to x}$, while the other direction $I_{x \to y}$ is largely unaffected by this weight over three different magnitudes of coupling strength. However, for couplings smaller than $~10^-3$ no significant causal influence is detectable. 
We varied the coupling for different amounts of data (Fig. \ref{fig:figure4} \textbf{(a)}, \textbf{(d)}). For this system at least $10^3$ data points are needed to detect significant causal influences. If the coupling is small the required amount of data increases.  Note that, regardless of the amount of data, no false positives are detected. To demonstrate noise robustness we injected intrinsic and external additive Gaussian noise in the logistic maps FIG. \ref{fig:figure4} \textbf{(c) - (f)}. While noise lowers the causal influence in both cases, it still correctly depends on the coupling and even for strong noise no false positives are introduced.

\begin{figure}[h!]
\includegraphics[width=.5\textwidth]{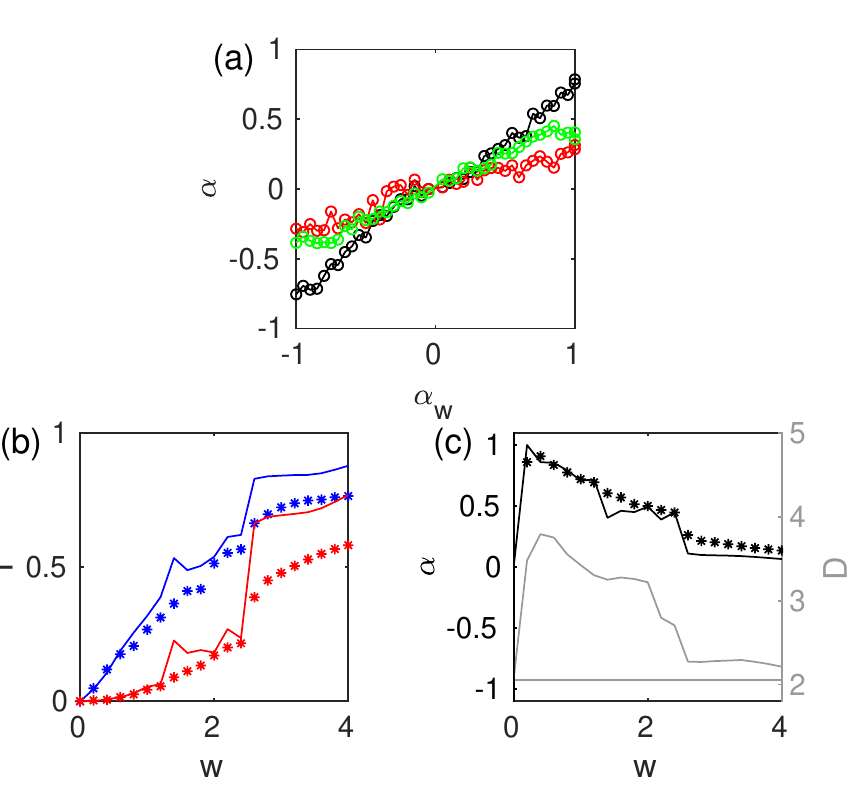}
\centering
\caption{\textbf{(a)} Asymmetry index between two Lorenz oscillators ($N=10^4$ data points) with slightly different frequencies and coupled by their $y$-components ($\theta_{1/2}=10, \rho_1=28.5, \rho_2=27.5, \theta_{1/2} = 8/3$). The time series were embedded with $m=9$ and $\tau=10$ and $E=10^2$ were used to estimate chance-level. The noise free asymmetry is shown in black. Coloured circles represent internal noise with $\sigma=2$ (green) and additive external noise with $\sigma=2$ (red). 
\textbf{(b),(c)} Unilaterally coupled Roessler (X) $\rightarrow$ Lorenz (Y) System. $10^3$ reference points were chosen from $10^4$ data points with $\tau=2$ and $m=7$. For chance-level erstimation $E=10^3$ ensembles were used. \textbf{(b)} Causal Influence $I_{x \to y}$ (blue) and $I_{y \to x}$ (red) for the noise free (solid-lines) and  for the systems perturbed by Gaussian white noise (asterix's). \textbf{(c)} The asymmetry for the noise free (solid lines) and perturbed system (asterix's). Grey Lines show the  Information Dimension estimated from the respective time series.}
\label{fig:figure5}
\end{figure}

Next we demonstrate the correct identification of the direction of causal influence for time continuous systems using two coupled Lorenz-systems. Here, we additionally introduced external and internal noise to demonstrate noise resistance also in this case (Fig. \ref{fig:figure5}). The Lorenz systems used were coupled by their $x$-components and are given by:
\begin{align*}
& \dot{x}_i = -\mu_i (x_i -y_i)+w_{ij}x_j+ \sigma \eta^{x_i}\\
& \dot{y}_i = \rho_i x_i - y_i - x_i z_i+ \sigma \eta^{y_i} \\
& \dot{z}_i = -\theta_i z_i + x_iy_i+ \sigma \eta_i^{z_i} \\
& \text{with} \quad <\eta^i\eta^j>=\delta_{ij}\delta(t-t'), 
\end{align*}
with parameters $\mu_i = 28$, $\rho_i=8/3$ and $\theta_i=10$.
For both, the noise free and noise polluted system, the correct direction of causal influence is determined correctly by $\alpha$. However, in the presence of noise the causal influence is weakened and the asymmetry less pronounced, which was also observed for the logistic maps.

Finally, we investigate the causal influence for strong couplings, where systems tend to synchronize. To quantify synchronization we estimate the information dimension from $d_i^i(k)$. For unilateral coupling this is a sensitive measure, since for complete synchronization the dimension of the driven system will drop to the one of the driving system. As an example we use a Roessler-Lorenz-System as analysed in \citep{Krakovska_2016}. In this system synchronization occurs at a critical value of $w \approx 2.5$, where the dimensions of the reconstructions coincide. CPM is not only able to detect the correct direction of causal influence before, but also after the critical coupling is reached (Fig. \ref{fig:figure5} b),c)). This result is in sharp contrast to other approaches (Krakovsk\'a Fig.4 \citep{Krakovska_2016}). Further tests in other systems (e.g. two coupled Fitzhugh-Nagumo Neurons) showed similar results up to the critical value. Interestingly, we found that intrinsic noise improves the detectability of influence asymmetries (asterix's in Fig. \ref{fig:figure5} c)).

\section{Discussion:}

When a system's component X unilaterally influences another component Y, then Y recieves information about X. X will not have as large proportions of information about states of Y, since in this case the latter typically have independent degrees of freedom. This simple heuristics is somewhat opposite to standard approaches based on prediction in direction of the influence including Granger Causality \citep{Granger_1969} and Transfer Entropy \citep{Schreiber_2000}. Here, we consider the relative amounts of information about the systems' states in the directions opposite to the influences. The cross prediction method estimates metric inflations of neighbourhoods and their projections via the putatively homeomorphic mappings among manifolds of reconstructed system states as a proxy for information loss. In the past a related idea was used for determining the quality of mappings \citep{Liebert_1991} \citep{Bauer_1992}, where the ratio of the inflations \textit{within} the respective spaces was expected to be close to one for homeomorphy while for topology violations it systematically deviates from one. In other words, while the core of method presented here is information theoretic its particular sensitivity for metric topology violations could have been expected from previous work. A recent work \citep{Harnack_2017} similarly used local properties of the mappings among state reconstructions to establish a measure of causal influence (termed Topological Causality). In stark contrast to Topological Causality, however, CPM exploits expansions \textit{within} and not \textit{among} reconstructions. Basing the method on the relation of distances within the same space solves a range of problems of TC and related approaches including Convergent Cross Mapping \citep{Sugihara_2012}. In particular, CPM is much less sensitive to synchronizations where to our knowledge previous methods often deliver misleading results \citep{Deyle_2016}\citep{Baskerville_2017}. \\
We are confident that applications also to real data will provide more reliable results than previous methods. For example an application to heart rate versus breathing rate data from an apnea patient \cite{Mark_1991} revealed a substantial asymmetry of influence from heart rate to breathing rate (Fig. \ref{fig:figure2} c), d)), an unambiguous result that is more pronounced than when determined with Transfer Entropy from the same data (Fig. 4 in \citep{Schreiber_2000}). \\

We thank M. Sch\"unemann for helpful comments on the manuscript.

\bibliography{refs}

\end{document}